\documentstyle[11pt]{article} 

\def\r{\bf r}
\def\R{\bf R}

\large{
\centerline{\bf Constant of motion, Lagrangian and Hamiltonian of the gravitational}
\vskip1pc
\centerline{\bf attraction of two bodies with variable mass}
}
\centerline{ G. L\'opez}
\centerline{Departamento de F\'{\i}sica de la Universidad de Guadalajara}
\centerline{Apartado Postal 4-137}
\centerline{44410 Guadalajara, Jalisco, M\'exico}   
\centerline{PACS: 03.20.+i} 
\centerline{Octuber, 2005}
\begin{document}  
\large{
\centerline{\bf ABSTRACT}
} 
\vskip1cm\noindent
The Lagrangian, the Hamiltonian and the constant of motion of the gravitational attraction of two
bodies when one of them has variable mass is considered. The relative and center of mass
coordinates are not separated, and choosing the reference system in the body with much higher
mass, it is possible to reduce the system of equations to 1-D problem. Then, a constant of motion,
the Lagrangian, and the Hamiltonian are  obtained. The trajectories found in the space
position-velocity,($x,v$), are qualitatively different from those on the space
position-momentum,($x,p$).
\vfil\eject
\noindent
{\bf 1. Introduction}
\vskip1pc\noindent
Mass variable systems has been important since the foundation of the classical mechanics and have
been relevant in modern physics [1]. Among these type of systems one could mentioned: the motion
of rockets [2], the kinetic theory of dusty plasmas [3], propagation of electromagnetic waves in
dispersive and nonlinear media [4], neutrinos mass oscillations [5], black holes formation [6],
and comets interacting with solar wind [7]. The interest in this last system comes from the
concern about to determinate correctly the trajectory of the comet as its mass is changing. This
system belong to the so called two-bodies problem. The gravitational two-bodies system is one of
the must well known systems in classical mechanics [8] and is the system which made a revolution
in our planetary and cosmological concepts. Normally, one assumes that the masses of these two
bodies are fixed and unchanged during the dynamical interaction [9]. However, this can not be true
any more when one consider comets as one of the bodies. Comets loose part of their mass as traveling
around the sun (or other star) due to their interaction with the solar wind which blows off
particles from their surfaces. In fact, it is possible that the comet could disappear as it
approaches to the sun [10]. So, one should consider the problem  of having one body with variable
mass  during its gravitational interaction with other body. 
\vskip0.5pc\noindent
In this paper, one considers the problem of finding the constant of motion, Lagrangian, and
Hamiltonian, for the gravitational interaction of two bodies when one of them is loosing its mass
during the gravitational interaction. The mass of one of the bodies is assumed much larger than the
mass of the other body. Choosing the reference system on big-mass  body, the three-dimensional
two-bodies problem is reduced to a one-dimensional problem. Then, one uses the constant of motion
approach [11] to find the Lagrangian and the Hamiltonian of the system. A model for the mass
variation is given for an explicit illustration of form of these quantities. With this model, one
shows that the trajectories in the space position-velocity (defined by the constant of motion) are
different than the trajectories on the space position-momentum (defined by the Hamiltonian). 
\vfil\eject
\leftline{\bf 2. Reference system and constant of motion}
\vskip0.5pc\noindent
Newton's equations of motion for two bodies interacting gravitationally, seen from arbitrary
inertial reference system, are given by
$${d\over dt}\left(m_1{d{\r_1}\over dt}\right)=-{Gm_1m_2\over
|{\r_1}-{\r_2}|^3}(\r_1-\r_2)\eqno{(1a)}$$ and
$${d\over dt}\left(m_2{d{\r_2}\over dt}\right)=-{Gm_1m_2\over |{\r_2}-{\r_1}|^3}(\r_2-\r_1)\
,\eqno{(1a)}$$
where $m_1$ and $m_2$ are the masses of the bodies, ${\r_1}=(x_1,y_1,z_1)$ and \hfil\break
${\r_2}=(x_2,y_2,z_2)$ are the vectors position of the two bodies from our reference system, $G$ is
the gravitational constant, and
$${|\r_1-\r_2|=|\r_2-\r_1|}=\sqrt{(x_2-x_1)^2+(y_2-y_1)^2+(z_2-z_1)^2}$$ is the Euclidean distance
between the two bodies. It will be assumed that $m_1$ is constant and that
$m_2$ varies with respect the time. Taking into consideration this mass variation,  Eqs. (1a) and
(1b) are written as
$$m_1{d^2{\r_1}\over dt^2}=-{Gm_1m_2\over |{\r_1}-{\r_2}|^3}(\r_1-\r_2)\eqno{(2)}$$ 
and
$$m_2{d^2{\r_2}\over dt^2}=-{Gm_1m_2\over |{\r_2}-{\r_1}|^3}({\r_2-\r_1})-\dot{m}_2{d{\r_2}\over dt}
\ ,\eqno{(3)}$$
where it has been defined $\dot{m}_2$ as $\dot{m}_2=dm_2/dt$. Now, let us consider the usual
relative, $\r$, and center of mass, $\R$, coordinates defined as
$${\r}={\r_2}-{\r_1}\ ,\quad\quad\hbox{and}\quad\quad{\R}={m_1{\r_1}+m_2{\r_2}\over m_1+m_2}\
.\eqno{(4)}$$ Let us then differentiate twice these coordinates with respect the time, taking into
consideration the equations (2) and (3). Thus, the following equations are obtained
$$\ddot{\r}=-{(m_1+m_2)G\over r^3}~{\r}-{\dot{m}_2\over m_2}~\dot{\r}_2\eqno{(5)}$$
and
$$\ddot{\R}={-\dot{m}_2\over m_1+m_2}~\dot{\r}_2+{2m_2\dot{m}_2\over(m_1+m_2)^2}~\dot{\r}+
{(m_1+m_2)m_1\ddot{m}_2-2m_1\dot{m}_2^2\over(m_1+m_2)^3}~\r\ .\eqno{(6)}$$
One sees that the relative motion does not decouple from the center of mass motion. So, these new
coordinates are not really useful to deal with mass variation systems. In fact, using (4) , one
has
$${\r_2}={\R}+{m_1\over m_1+m_2}~{\r}\ ,\quad\hbox{and}\quad\dot{\r}_2=\dot{\R}+{m_1\over
m_1+m_2}\dot{\r}-{m_1\dot{m}_2\over (m_1+m_2)^2}{\r}\ .\eqno{(7)}$$
Substituting these expressions in (5) and (6), one can see more clearly this coupling,
$$\ddot{\r}=\biggl[{m_1+m_2)G\over r^3}+{m_1\dot{m}_2^2\over m_2(m_1+m_2)^2}\biggr]{\r}-
{\dot{m}_2\over m_2}\biggl[\dot{\R}+{m_1\over m_1+m_2}\dot{\r}\biggr]\eqno{(8)}$$
and
$$\ddot{\R}={-\dot{m}_2\over m_1+m_2}{\R}+{m_1\dot{m}_2\over(m_1+m_2)^2}\dot{\r}+
{(m_1+m_2)m_1\ddot{m}_2-m_1\dot{m}_2^2\over(m_1+m_2)^3}~{\r}\ .\eqno{(9)}$$
However, one can consider the case for $m_1\gg m_2$ (which is the case star-comet), and consider
to put our reference system just on the first body ($\r_1=\vec 0$). In this case, Eq. (3) 
becomes
$$m_2{d^2{\r}\over dt^2}=-{Gm_1m_2\over r^3}~{\r}-\dot{m}_2\dot{\r}\ ,\eqno{(10)}$$
where ${\r}={\r_2}=(x,y,z)$. Using spherical coordinates ($r,\theta,\varphi$),
$$x=r\sin\theta\cos\varphi\ ,\ \ y=r\sin\theta\sin\varphi\ ,\ \ z=r\cos\theta\ ,\eqno{(11)}$$
Eq. (10) can be written as
$$m_2{d^2{\r}\over dt^2}=-\biggl[{Gm_1m_2\over r^2}+\dot{m}_2\dot{r}\biggr]~\widehat{r}+
\dot{m}_2\biggl(r\dot{\theta}~\widehat{\theta}+r\dot{\varphi}\sin\theta~\widehat{\varphi}\biggr)\
,\eqno{(12)}$$
where $\widehat{r}$, $\widehat{\theta}$ and $\widehat{\varphi}$ are unitary directional vectors,
$$\widehat{r}=(\sin\theta\cos\varphi, \sin\theta\sin\varphi,\cos\theta)\ ,\quad\hbox{with}\quad
\dot{\widehat{r}}=\dot{\theta}~\widehat{\theta}+\dot{\varphi}\sin\theta~\widehat{\varphi}\ ,\
\eqno{(13a)}$$
$$\widehat{\theta}=(\cos\theta\cos\varphi, \cos\theta\sin\varphi, -\sin\theta)\ ,\quad\hbox{with}
\quad\dot{\widehat{\theta}}=-\dot{\theta}~\widehat{r}+\dot{\varphi}\cos\theta~\widehat{\varphi}
\eqno{(13b)}$$
and
$$\widehat{\varphi}=(-\sin\theta, \cos\theta, 0)\ ,\quad\quad\hbox{with}\quad\quad
\dot{\widehat{\varphi}}=\sin\theta~\widehat{r}+\cos\theta~\widehat{\theta}\ .\eqno{(13c)}$$
Since one has that ${\r}=r\widehat{r}$, it follows that
\begin{eqnarray*}
\ddot{\r}&=&(\ddot{r}-r\dot{\theta}^2+r\dot{\varphi}\sin^2\varphi)\widehat{r}+
(2\dot{r}\dot{\theta}+r\ddot{\theta}+r\dot{\varphi}\sin\theta\cos\theta)\widehat{\theta}\\
& &+
(2\dot{r}\dot{\varphi}\sin\theta+r\ddot{\varphi}\sin\theta+2\dot{\varphi}\dot{\theta}\cos\theta)
\widehat{\varphi}\ ,\hskip5cm (14)
\end{eqnarray*}
and Eq. (12) is discomposed in the following three equations
$$m_2(\ddot{r}-r\dot{\theta}^2+r\dot{\varphi}\sin^2\varphi)=-{Gm_1m_2\over r^2}-\dot{m}_2\dot{r}\
,\eqno{(15a)}$$
$$m_2(2\dot{r}\dot{\theta}+r\ddot{\theta}+r\dot{\varphi}\sin\theta\cos\theta)=
\dot{m}_2r\dot{\theta}\ ,\eqno{(15b)}$$
and
$$m_2(2\dot{r}\dot{\varphi}\sin\theta+r\ddot{\varphi}\sin\theta+2\dot{\varphi}\dot{\theta}\cos\theta)
=\dot{m}_2r\dot{\varphi}\sin\theta\ .\eqno{(15c)}$$
Thus, one has obtained coupling among these coordinates due to the term $\dot{m}_2$. Nevertheless,
one can restrict oneself to consider the case $\dot{m}_2r\approx 0$. For this case, it follows that
$\dot{\varphi}=0$, and the resulting equations are
\vfil\eject
$$m_2(\ddot{r}-r\dot{\theta}^2)=-{Gm_1m_2\over r^2}-\dot{m}_2\dot{r}\ ,\eqno{(16a)}$$
and
$$m_2(2\dot{r}\dot{\theta}+r\ddot{\theta})=0\ .\eqno{(16b)}$$
Let $m_o$ be the mass of the second body when this one is very far away from the first body (when
a comet is very far away from the sun, the mass of the comet remains constant). Since $m_2\not=0$
on (16b), the expression inside the parenthesis must be zero. In addition, one can multiply this
expression by $m_o r$ to get the following constant of motion
$$P_{\theta}=m_or^2\dot{\theta}\ .\eqno{(17)}$$
Using this constant of motion in (16a), one obtains the equation
$${d^2r\over dt^2}=-{Gm_1\over r^2}+{P_{\theta}^2\over m_o^2r^3}
-{\dot{m}_2\over m_2}\left({dr\over dt}\right)\ .\eqno{(18)}$$
This equation represents a dissipative system for $\dot{m}_2>0$ and 
anti-dissipative system for $\dot{m}_2<0$. Suppose now that $m_2$ is a function of the distance
between the  first and second body, $m_2=m_2(r)$. Therefore, it follows that
$${dm_2\over dt}={dm_2\over dr}{dr\over dt}\ ,\eqno{(19)}$$
and Eq. (18) can be written as
$${d^2r\over dt^2}=-{Gm_1\over r^2}+{P_{\theta}^2\over m_o^2r^3}
-{{m}_2'\over m_2}\left({dr\over dt}\right)^2\ ,\eqno{(20)}$$ 
where $m_2'=dm_2/dr$. 
This equation can be seen as the following autonomous dynamical system [12]
$${dr\over dt}=v\ ,\quad\quad{dv\over dt}=-{Gm_1\over r^2}+{P_{\theta}^2\over m_o^2r^3}
-{{m}_2'\over m_2}v^2\ .\eqno{(21)}$$ 
A constant of motion for this system is a function $K=K(r,v)$ such that the following partial
differential equation is satisfied [13]
$$v{\partial K\over\partial r}+\biggl({-Gm_1\over r^2}+{P_{\theta}^2\over m_o^2r^3}
-{m_2'\over m_2}v^2\biggr){\partial K\over\partial v}=0\ .\eqno{(22)}$$
This equation can be solved by the characteristic method [14] from which the following
characteristic curve results
$$C(r,v)=m_2^2(r)v^2+2Gm_1\int{m_2^2(r)~dr\over r^2}-{2P_{\theta}^2\over
m_o^2}\int{m_2^2(r)~dr\over r^3}\ ,\eqno{(23)}$$
and the general solution of (22) is given by
$$K(r,v)=F(C(r,v))\ ,\eqno{(24)}$$
where $F$ is an arbitrary function of the characteristic curve. One can have a constant of motion
with units of energy by selecting $F$ as $F=C/2m_o$. That is, the constant of motion is given by
$$K(r,v)={m_2^2(r)\over 2m_o}v^2+{Gm_1\over m_o}\int {m_2^2(r)~dr\over r^2}-
{P_{\theta}^2\over m_o^3}\int{m_2^2(r)~dr\over r^3}\ .\eqno{(25)}$$
\vskip1pc\noindent
\leftline{\bf 2. Lagrangian and Hamiltonian}
\vskip0.5pc\noindent
Given the time independent constant of motion (25), the Lagrangian of the system (20) can be
obtained using the following known expression [11]
$$L(r,v)=v\int{K(r,v)~dv\over v^2}\ .\eqno{(26)}$$
Thus, the Lagrangian is given by
$$L(r,v)={m_2^2(r)\over 2m_o}v^2-{Gm_1\over m_o}\int {m_2^2(r)~dr\over r^2}+
{P_{\theta}^2\over m_o^3}\int{m_2^2(r)~dr\over r^3}\ .\eqno{(27)}$$
The generalized linear momentum ($p=\partial L/\partial v$) is 
$$p={m_2^2(r)\over m_o}v\ ,\eqno{(28)}$$
and the Hamiltonian is
$$H(r,p)={m_o p^2\over 2m_2^2(r)}+{Gm_1\over m_o}\int {m_2^2(r)~dr\over r^2}-
{P_{\theta}^2\over m_o^3}\int{m_2^2(r)~dr\over r^3}\ .\eqno{(29)}$$
Note from (25) and (29) that the constant of motion and Hamiltonian can be written as
$$K(r,v)={m_2^2(r)\over 2m_o}v^2+V_{eff}(r)\eqno{(30)}$$
and
$$H(r,p)={m_o p^2\over 2m_2^2(r)}+V_{eff}(r)\ ,\eqno{(31)}$$
where $V_{eff}$ is the effective potential energy defined as
$$V_{eff}(r)={Gm_1\over m_o}\int {m_2^2(r)~dr\over r^2}-
{P_{\theta}^2\over m_o^3}\int{m_2^2(r)~dr\over r^3}\ .\eqno{(32a)}$$
This potential energy has an extreme value at the point
$$r^*={P_{\theta}^2\over Gm_1m_o^2}\eqno{(32b)}$$
which depends on $m_o$ but it does not depend on the model for $m_2(r)$. One can see that this
extreme value is a minimum for $m_2(r^*)\not=0$, since one has that
$$\left({d^2V_{eff}\over dr^2}\right)_{r=r^*}={(Gm_1m_o)^4m_om_2^2(r^*)\over P_{\theta}^6}>0\ .$$ 
On the other hand, because of
the expression (28), one could expect different behavior of a trajectory in the phase space ($r,v$)
and the phase space ($r,p$). The trajectory $r(\theta)$  is found using the relation
$dr/dt=(dr/d\theta)\dot{\theta}$, and the Eq. (17) in (30) to get
$$\int_{\theta_o}^{\theta}d\theta={P_{\theta}\over \sqrt{2m_o^3}}\int_{r_o}^r{m_2(r)~dr\over
r^2\sqrt{K-V_{eff}(r)}}\ ,\eqno{(34a)}$$
where $K$ and $P_{\theta}$ are determinate by the initial conditions, $K=K(r_o,v_o)$ and
$P_{\theta}=m_or_o^2\dot{\theta}_o$. The time of half of cycle of oscillation, $T_{1/2}$, is
obtained directly from Eq. (30) as
$$T_{1/2}={1\over \sqrt{2m_o}}\int_{r_1}^{r_2}{m_2(r)~dr\over\sqrt{K-V_{eff}(r)}}\ ,\eqno{(34b)}$$
where $r_1$ and $r_2$ are the two return points deduced as the solution of the following equation 
$$V_{eff}(r_i)=K\ ,\quad i=1,2\ .\eqno{(34c)}$$
\vskip2pc\noindent
\leftline{\bf 3. Model of Variable Mass}
\vskip0.5pc\noindent
As a possible application of (25) and (29), consider that a comet looses material as a result of
the interaction with star wind in the following way (for one cycle of oscillation)
$$m_2(r)=\cases{m_{oo}\sqrt{1-e^{-\alpha r}}& incoming ($v<0$)\cr\cr
m_ie^{\alpha(r_1-r)}+m_f(1-e^{-\alpha r})& outgoing ($v>0$)\cr}\eqno{(35)}$$
where $m_{oo}$ or $m_f$ (where $m_f=2m_i-m_{oo}$ by symmetry) is the mass of the comet very far
away from the star (in each case),
$m_i$ is the mass of the comet at the closets approach to the star (distance $r_1$),
$m_i=m_{oo}\sqrt{1-e^{-\alpha r_1}}$, and $\alpha$ is a factor that can be adjusted from
experimental data. Thus, the effective potential (32a) has the following form for the incoming case
($m_o=m_{oo}$)
\begin{eqnarray*}
V_{eff}^{(in)}(r)&=&-{Gm_1m_{oo}\over r}(1-e^{-\alpha r})+{P_{\theta}^2\over
2m_{oo}r^2}(1-e^{-\alpha r})\\
& & +\biggl[GM_1m_{oo}\alpha+{\alpha^2P_{\theta}^2\over
2m_{oo}}\biggr]Ei(-\alpha r)+ {\alpha P_{\theta}^2e^{-\alpha r}\over 2m_{oo}r}\ ,\hskip2.5cm(36a)
\end{eqnarray*}
where $Ei(x)$ is the exponential-integral function [15]. 
\vfil\eject\noindent
For the outgoing case, one has $m_o=m_f$ and
\begin{eqnarray*}
V_{eff}^{(out)}(r)&=&-{Gm_1m_f\over r}+{\tilde{P}_{\theta}^2\over 2m_f r^2}\\
& &+{Gm_1(m_ie^{\alpha r_1}-m_f)^2\over m_f}\biggl[-{e^{-2\alpha r}\over r}-2\alpha Ei(-2\alpha r)
\biggr]\\
& &-{\tilde{P}_{\theta}^2(m_ie^{\alpha r_1}-m_f)^2\over m_f^3}
\biggl[-{e^{-2\alpha r}\over 2r^2}+{\alpha e^{-2\alpha r}\over r}+2Ei(-2\alpha r)\biggr]\\
& &+2Gm_1(m_ie^{\alpha r_1}-m_f)\biggl[-{e^{-\alpha r}\over r}-\alpha Ei(-\alpha r)\biggr]\\
& &-{2\tilde{P}_{\theta}^2(m_ie^{\alpha r_1}-m_f)\over m_f^2}
\biggl[-{e^{-\alpha r}\over 2r^2}+{\alpha e^{-\alpha r}\over 2r}+{\alpha^2\over 2}Ei(-\alpha
r)\biggr]\ ,\hskip1cm(36b)
\end{eqnarray*}
where $\tilde{P}_{\theta}$ is defined now as $\tilde{P}_{\theta}=m_fr^2\dot{\theta}$. The extreme
point of the effective potential (32b) for the incoming and outgoing cases is given by
$$r^*_{in}={P_{\theta}^2\over Gm_1m_{oo}^2}\ ,\quad\quad r^*_{out}={P_{\theta}^2\over Gm_1m_f^2}\
.\eqno{(37)}$$
Given the definition (35), the constant of motion, Lagrangian, generalized linear momentum, and
Hamiltonian are given by
$$K^{(i)}(r,v)={m_2^2(r)\over 2m_o}v^2+V_{eff}^{(i)}(r)\ ,\eqno{(38)}$$
$$L^{(i)}(r,v)={m_2^2(r)\over 2m_o}v^2-V_{eff}^{(i)}(r)\ ,\eqno{(39)}$$
$$p^{(i)}(r,v)={m_2^2(r)\over m_o}v\ ,\eqno{(40)}$$
and
$$H^{(i)}(r,p)={m_o p^2\over 2m_2^2(r)}+V_{eff}^{(i)}(r)\ ,\eqno{(41)}$$
where $i=in$ for the incoming case, and $i=out$ for the outgoing case.  As an example of
illustration of this model, let us use the following parameters to estimate the dependence of
several physical quantities with respect the parameter $\alpha$,
\begin{eqnarray*}
G&=&6.67\times 10^{-11}m^3/Kg~sec\ ;\quad m_{oo}=10^6Kg\ ;\quad m_f=0.1~m_{oo}\ ;\\
P_{\theta}&=&10^{17}Kg~m^2/sec\ ;\quad\hbox{and}\quad K=-8\times 10^{23}Joules\ .\hskip2.8cm(42)
\end{eqnarray*}
Fig. 1 shows the curves of $V_{eff}(r)$ for several values of $\alpha$ (incoming case). As one
can see from this figure, the location of the minimum does not change, but the minimum value of
$V_{eff}$ tends to disappear as $\alpha$ goes to zero. Also for the incoming case, Fig. 2 shows how
the minimum distance of approximation of the two bodies, $r_1$, and maximum distance, $r_2$, behave
as a function of the parameter $\alpha$. As one can guess, the following limit is satisfied
$\lim_{\alpha\to 0}r_1=\lim_{\alpha\to 0}r_2=r^*$ which will become a inflexion point for
$V_{eff}$. Fig. 3 shows the velocity ($v$) and normalized linear momentum
($p/m_{o}$) as a function of $r$ for several values of $\alpha$ and for the incoming case. All the
trajectories start at $r_2=200$ and finish at $r_1(\alpha)$. One can see the difference of the
trajectories in (a) with respect to (b) due to position dependence of the momentum, relation (40).
\vskip2pc\noindent
\leftline{\bf Conclusions}
\vskip0.5pc\noindent
The Lagrangian, Hamiltonian and a constant of motion of the gravitational attraction of two bodies
when one of them has variable mass were given. One found that the relative and center of mass
coordinates are coupled due to this mass variation. However, chosen the reference system in the
much more massive body, it was possible to reduce the system to 1-D problem. Then, the constant of
motion, Lagrangian and Hamiltonian were obtained. One main feature of these quantities was the
appearance of an effective potential, which is reduced (when $\dot{m}_2=0$) to the usual
gravitational effective potential of two bodies with fixed masses. Other feature was the distance
dependence of the generalized linear momentum. A model for comet-mass-variation was given which
depends on the parameter $\alpha$. A study was made of the dependence with respect to $\alpha$ of
$V_{eff}$, minimum and maximum distance between the two bodies,  and the trajectories in the
spaces ($r,v$) and ($r,p$). Of course, the problem of the interaction comet-star with the variation
of mass deserves more complete analysis. The intention here with this example was to show explicitly
the form of the constant of motion, Lagrangian, and Hamiltonian and to point out the different
trajectories behavior in the spaces ($r,v$) and ($r,p$) arising from the constant of motion and
Hamiltonian.

\vfil\eject
\leftline{\bf References}
\vskip1pc\noindent
{\obeylines
1. G. L\'opez, L.A. Barrera, Y. Garibo, H. Hern\'andez, J.C. Salazar,
\quad and C.A. Vargas, Int. Jour. Theo. Phys.,{\bf 43},10 (2004),1.
2. A. Sommerfeld,{\it Lectures on Theoretical Physics}, Vol. I, 
\quad Academic Press (1964).
3. A.G. Zagorodny, P.P.J.M. Schram, and S.A. Trigger, Phys. Rev. Lett.,
\quad {\bf 84} (2000),3594.
4. O.T. Serimaa, J. Javanainen, and S. Varr\'o, Phys. Rev. A, 
\quad{\bf 33}, (1986), 2913. 
5. H.A. Bethe, Phys. Rev. Lett.,{\bf 56}, (1986),1305.
\quad E.D. Commins and P.H. Bucksbaum, {\it Weak Interactions of Leptons 
\quad and Quarks}, Cambridge University Press (1983).
6. F.W. Helhl, C. Kiefer and R.J.K. Metzler,{\it Black Holes: Theory and
\quad  Observation}, Springer-Verlag (1998).
7. J.A. Nuth III, H.G.M. Hill, and G. Kletetschka, Nature {\bf 406},(2000) 275.
\quad H. Reeves, Nature {\bf 248}, (1974) 398.
\quad L. Biermann, Nature {\bf 230}, (1971) 156.
8. H. Goldstein,{\it Classical Mechanics}, Addison-Wesley, M.A., (1950).
9. J.J. Matese, P.G. Whitman, and D.P. Whitmire, Nature {\bf 352}, 
\quad (1991) 506.
\quad J. Heisler, Nature {\bf 324}, (1986) 306.
10. S.A. Stern and P.A. Weissman, Nature {\bf 409}, (2001) 589.
\quad D.W. Hughes, Nature {\bf 308}, (1984) 16.
11. J.A. Kobussen, Acta Phys. Austr. {\bf 51},(1979),193.
\quad C. Leubner, Phys. Lett. A {\bf 86},(1981), 2.
\quad G. L\'opez, Ann. of  Phys., {\bf 251},2 (1996),372.
12. P.G. Drazin, {\it Nonlinear Systems}, Cambridge University Press, (1992),
\quad\quad  chapter 5.
13. G. L\'opez, Ann. of Phys., {\bf 251},2 (1996),363.
14. F. John,{\it Partial Differential Equations}, Springer-Verlag N.Y. (1974).
15. I.S. Gradshteyn and I.M. Ryzhik, {\it Table of Integrals, Series and
\quad Products}, Academic Press 1980, page 93.} 

\vfil\eject
\leftline{\bf Figure Captions}
\vskip0.5pc\noindent
Fig. 1 $V^{(in)}_{eff}(r)$ with the values of the parameters given on (42), for $\alpha=1$ (1);
$\alpha=0.01$ (2); and $\alpha=0.005$ (3).
\vskip1pc\noindent
Fig. 2 Maximum ($r_2$) and minimum ($r_1$) distances between the two bodies  as a function of the
parameter $\alpha$.
\vskip1pc\noindent
Fig. 3 (a): Trajectories in the plane ($r,v$); (b): Trajectories in the plane ($r,p$). $\alpha=1$
(1), $\alpha=0.01$ (2), and $\alpha=0.005$ (3).

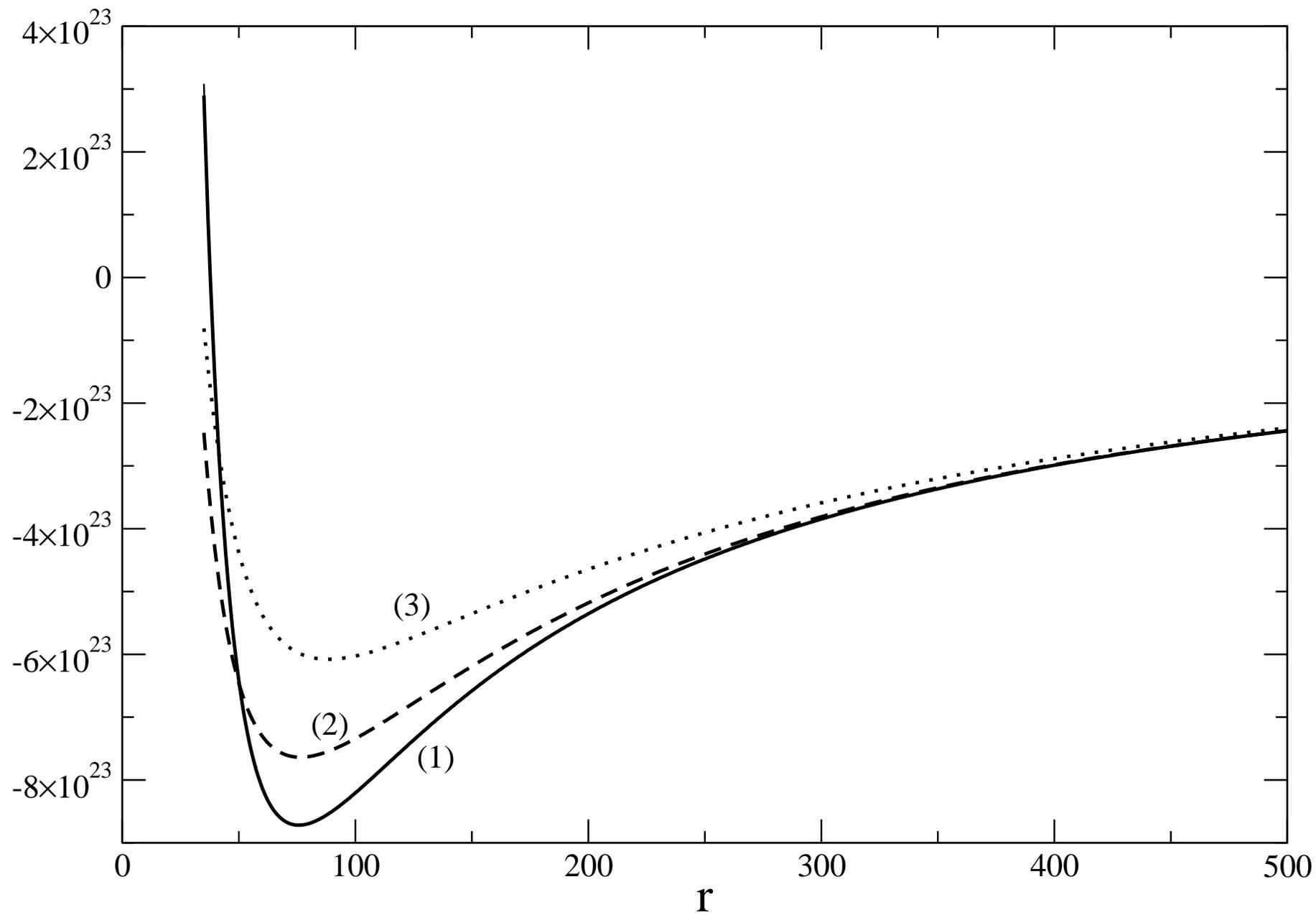
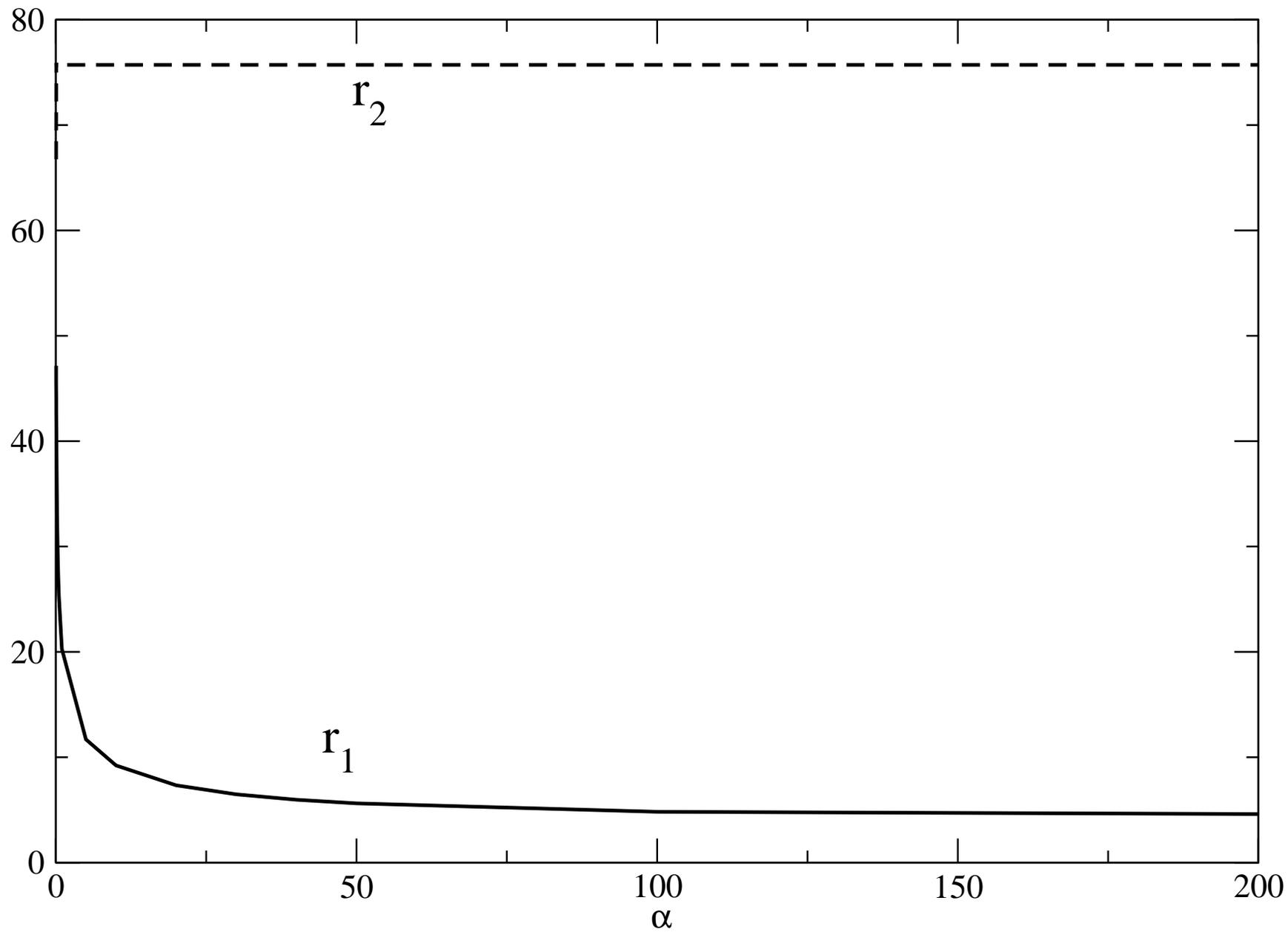
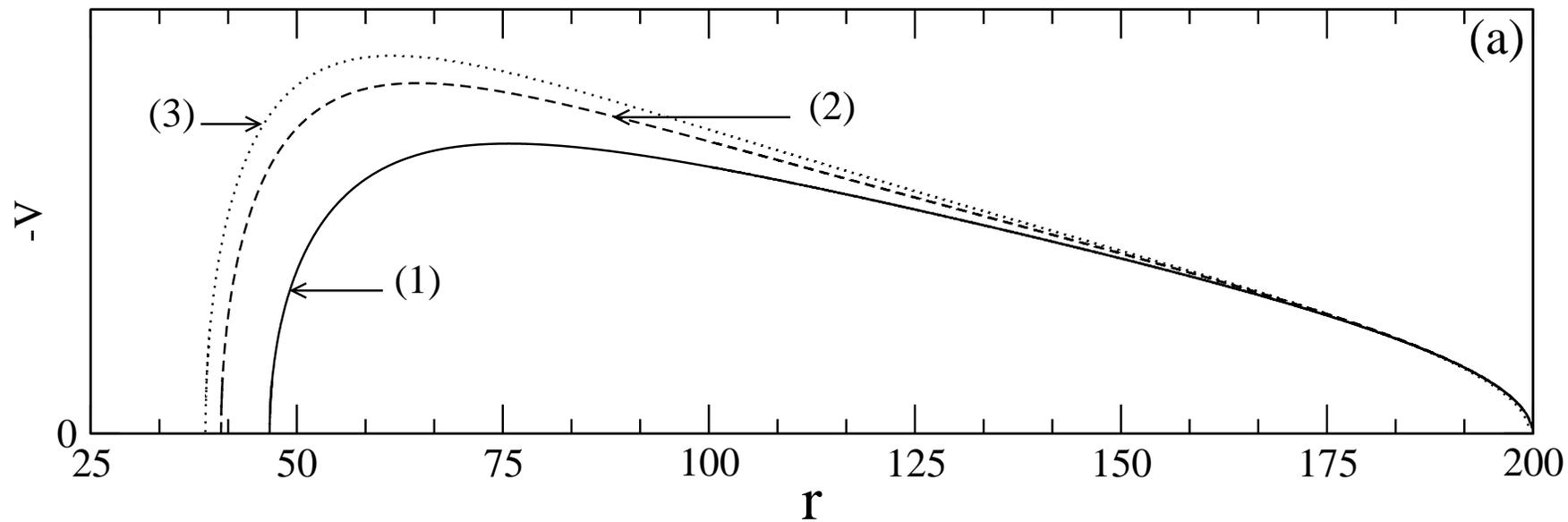
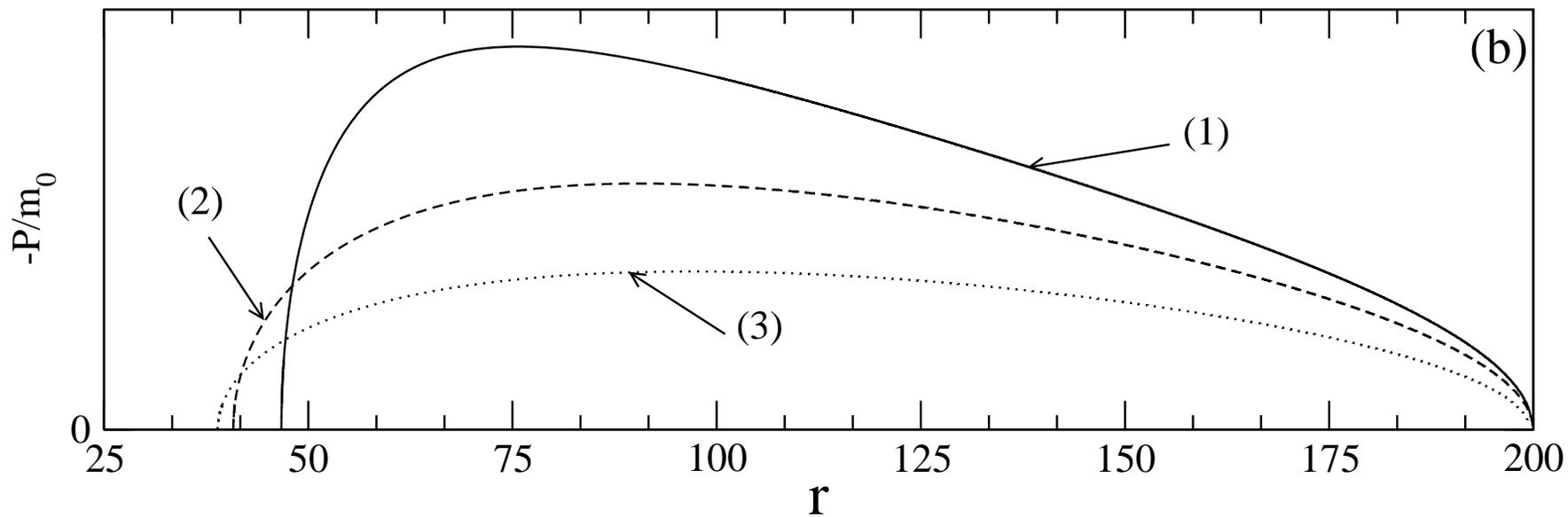
\end{document}